\documentclass[twocolumn]{aastex631}

\newcommand{\bpmg}{{$\beta$PMG }}
\usepackage{amsmath}

\shorttitle{Flares and the $\beta$ Pictoris Moving Group}

\begin{document}
\title{Flaring Activity for Low-Mass Stars in the $\beta$ Pictoris Moving Group}

\author[0000-0002-2764-1217]{Jordan N. Ealy}
\affiliation{Department of Astronomy, University of Maryland, College Park, MD 20782, USA}
\affiliation{NASA Goddard Space Flight Center, Greenbelt, MD 20771, USA}

\author[0000-0001-5347-7062]{Joshua E. Schlieder}
\affiliation{NASA Goddard Space Flight Center, Greenbelt, MD 20771, USA}

\author[0000-0002-9258-5311]{Thaddeus D. Komacek}
\affiliation{Department of Astronomy, University of Maryland, College Park, MD 20782, USA}

\author[0000-0002-0388-8004]{Emily A. Gilbert}
\affiliation{Jet Propulsion Laboratory, California Institute of Technology, 4800 Oak Grove Drive, Pasadena, CA 91109, USA}
\begin{abstract}
Stellar flares from K and M dwarfs release panchromatic radiation characterized by a significantly higher brightness temperature ($\sim$9-20 kK) than the star. The increased frequency of magnetic activity on young low-mass stars results in the energy released during flaring events becoming a notable contributor to the radiation environment.  This study focuses on the $\beta$ Pictoris moving group (24 $\pm$ 3 Myr) for the analysis of young, low-mass star flaring rates within the framework of larger flare studies. The calibration of long-term optical flare statistics is crucial to updating flare activity-age relations and the interpretation of exoplanet atmosphere observations. Using the $\beta$ Pictoris moving group, we develop a modular flare extraction pipeline sensitive to low-mass stellar flares in observations from the \textit{Transiting Exoplanet Survey Satellite}. This pipeline is built to characterize flare properties of these stars such as total energy and cumulative flare rate. Consistent with previous studies, this sample (N=49) shows higher cumulative flare rates than early type and old field stars by at least an order of magnitude. Fitted flare frequency distributions for both early and late type M dwarfs show an average slope of  $1.58\, \pm \, 0.23$ with earlier stars flaring with lower or similar rates to late types. A typical member in this sample has daily ($\mathrm{\sim 1 \, d^{-1}}$ ) flares with TESS band energies of  $10^{32} - 10^{33}$ ergs. The optical flare rates and energies for this group provide essential context into the co-evolution of host stars and associated planets.

\end{abstract}


\section{Introduction} \label{sec:intro}
Representing $\sim$70\% of the stellar population, M dwarfs’ relatively cool temperatures and small radii make these targets advantageous for planet detection and atmospheric characterization compared to earlier type main-sequence stars \citep{Henry2006}. The increased likelihood of transiting planets due to high occurrence rates for short period ($<100$ d) terrestrial planets ($\sim 1$ per star) around M dwarfs \citep[][]{DC+13,DC+15,HD+2019,S+21} make these systems outstanding candidates for exoplanetary characterization. M dwarf planetary systems such as AU Mic \citep[][]{Pl20,M+21,G+22}, TRAPPIST-1 \citep[][]{G+17}, GJ 1002 \citep[][]{suz2023}, and Proxima Centauri \citep[][]{A+16} have already offered a unique look at the diversity of exoplanet systems in the solar neighborhood.

Both empirical and statistical models have been developed to independently characterize stellar and planetary properties for exoplanetary systems \citep[e.g.,][]{S+07,Mann2015,S+21} with studies focusing on exploring the interdependence of planetary evolution and the physical attributes of their host stars becoming more prevalent \citep[][]{F+05,S+13,Kan+19}.

\citet{Y+14} explored the relationship between system architecture and the habitable zone with 3D global climate models. Due to the short orbital separations, planets orbiting M dwarfs are more likely to be tidally locked. This effect, along with enhanced absorption of low energy stellar photons by water vapor and other greenhouse gases in the atmosphere, results in a cloudier dayside of the planet. The increased albedo due to cloud cover results in an extended inner edge of the habitable zone \citep[][]{Kop+13}. 

Particularly for low mass stars, effects due to magnetic activity play an important role. The enhanced magnetic activity of M dwarfs compared to higher-mass stars produce observable phenomena such as flaring and star spots. Stellar flares occur as a result of magnetic reconnection \citep[][]{Martens89}. These high energy ($>10^{31}$ ergs  in the \textit{TESS} band; \citet{G2020}) events release panchromatic radiation and have higher blackbody temperatures than M dwarf photospheres.

This emission is usually approximated by a hot (9-10 kK) blackbody despite various physical sources and emitting mechanisms with the majority of the emission being in the near-ultraviolet (UV) and optical \citep[][]{OW2015,howard18,jack18}. Due to the enhanced magnetic activity of M dwarfs compared to higher mass stars, the energy released (ex. flares) is a non-negligible contributor to the overall radiation environment near the star \citep{SB2014}.

The augmented UV emission can significantly impact observables such as photochemically-induced spectral features in transmission and atmospheric erosion \citep[e.g.,][]{K+19,chen21,doAm+22}. Recent spectroscopy of flares has shown that the spectral energy distribution is more complex, requiring nuanced investigation into the timescales and relative importance of salient processes that can impact observations  \citep[][]{Loyd16,Y2017,mac21}. 

Studies that focus on the impact of stellar magnetic activity add context to current evolution models and observational predictions.
Historically, this has taken many forms including effects from single flares, continuous activity, and laboratory work \citep[e.g.,][]{segura2010,T2019, doAm+22,ranjan+17,Rimmer2018,Abre2020}. The effects of flares can extend to the destruction of ozone layers necessary to insulate the surface from UV \citep[][]{T2019,Young+17}, desiccation \citep[][]{doAm+22}, and the sterilization of planetary surfaces \citep[][]{G2020,Spinelli2023}. In contrast, \citet{ranjan+17} present flares as a necessary mechanism for the generation of UV photons which can initiate prebiotic chemistry for low-mass stars. Recently, \citet{louca+22} simulated planetary atmospheres of different compositions exposed to stellar activity from low-mass stars. They showed that for both hydrogen- and nitrogen-dominated atmospheres, continued exposure to the high energy radiation from flares can both gradually and abruptly alter the composition in the observationally accessible portions of the atmosphere. In addition to this result, the continued flaring activity within their models photochemically produced conditions that sustained permanent compositional changes. As a result, determining flare properties is essential to fully characterizing exoplanet hosts, their prospects for habitability, and the co-evolution of stars and their planets using both models and observations.

\begin{figure*}
	\includegraphics[width=2\columnwidth]{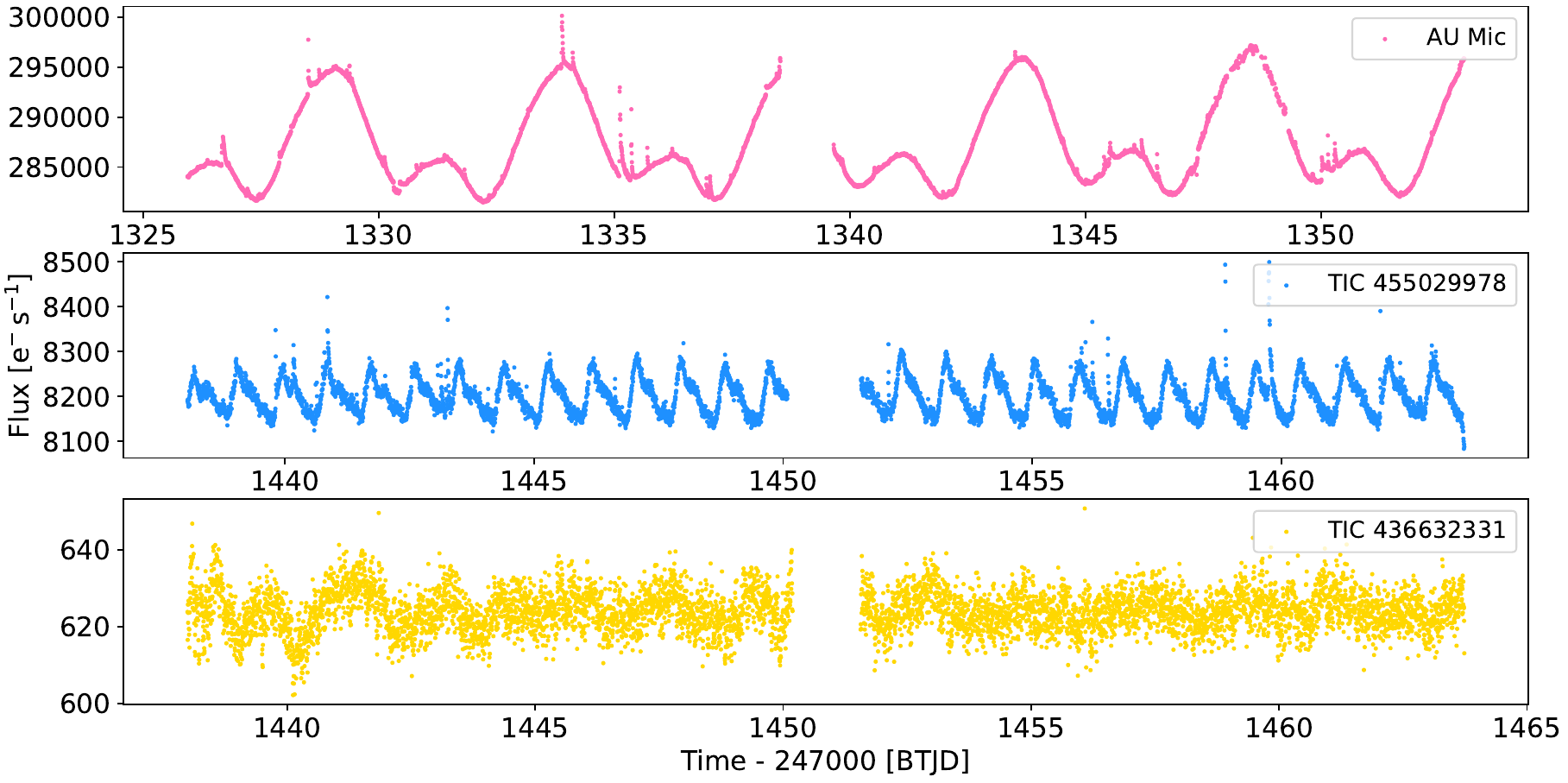}
    \caption{Due to sustained star spot coverage and fast rotation periods of active low-mass stars, their light curves have distinct patterns. 
   The sharp peaks superimposed on the spot modulation are flares. The varied amplitude, duration, and temporal distribution offer insight into the mechanisms generating them.}
    \label{fig:spotmod}
\end{figure*}

Large scale flare studies have been critical in establishing and testing flare models \citep[][]{dave14,Loyd16}, constraining bolometric flare rates for various spectral types \citep[][]{G2020,Ilin21}, and conducting multiwavelength studies to determine spectral energy partitions (\citet{mac21,paudel21,Paudel+2024}; Vega et al. in prep.).  At the same time, variations in metallicity, activity cycles, and particularly age become less pronounced due to population binning. \citet{Ilin2019} and \citet{dave+19} have shown that young stars are more magnetically active than older main sequence stars by comparing flaring rates using \textit{Kepler}. In addition, M dwarf magnetic field structure varies due to the M4 partial to fully convective boundary \citep[][]{k11}. Both conditions suggest notable evolution of young, late-type stars during early epochs of planetary development. As low stellar mass planetary system detections and confirmations continue, in-depth analyses of stellar activity with system ages will provide a more accurate picture of the evolution of the total radiation output, including flares.

With main-sequence lifetimes longer than the current age of the universe, the age of a random field M dwarf is difficult to precisely determine. In addition to this, these stars are intrinsically dim, making them difficult to characterize fully without significant integration times or large-aperture instruments. Both of these challenges can be remediated through the use of moving groups. These groupings include coeval stars that were born from the same molecular cloud complex and retain common galactic motion. Since moving groups' demographics are determined by the initial mass function, determining the ages of more massive members of these associations using methods like gyrochronlogy or evolution model fits are more accessible than dating the low-mass population directly. Due to these constraints, it is more efficient to date the more massive members to establish the bulk properties of the group, then identify kinematically related lower-mass stars and verify consistent ages through activity, lithium depletion, low-gravity spectral features, and other age diagnostics \citep{banyan}. This allows for well-calibrated ages accurate to within a few million years \citep[][]{Kiman2021}.

\citet{banyan} present 27 recognized nearby young associations and moving groups in the Milky Way with ages ranging from 10 Myr to 750 Myr. This range of ages covers stellar and planetary evolutionary stages including transitions to the main sequence, planetary accretion \& migration, and early atmospheric evolution. For this study, we focus on the $\beta$ Pictoris moving group ($\beta$PMG). At an average distance of 60 pc, the M dwarfs of this group are bright and spread across the sky. They are also at the pivotal age of $24 \pm 3$ Myr where planets have formed and are undergoing early evolution \citep[][]{bell15}. The M1 exoplanet host and $\beta$PMG member, AU Mic, currently has two confirmed Neptune-sized planets orbiting interior to a debris disk \citep[][]{M+21,G+22}.
With $>100$ confirmed and prospective low mass members, this moving group provides an ideal launch point for analyzing flare rates of young low mass stars.

Our methods for stellar sample constraints and the development of a modular flare pipeline are discussed in Section \ref{sec:methods}. A summary of $\beta$PMG flare properties both intra-group and compared to large flare studies are presented in Section \ref{sec:results}. Broader impacts to local stellar environments, star-planet interactions, and physical implications of age-dependent variations of flaring activity are summarized in Section \ref{sec:cons/disc}.


\section{Methods} \label{sec:methods}
\subsection{Determining the $\beta$ Pictoris Low Mass Stellar Sample}\label{sec:B182}

To refine a conservative list of $\beta$ Pictoris members, we use previous literature surveys, new kinematic data, and membership probability software to constrain high likelihood candidates.
The initial census for this study was generated by \cite{S2017} through an astrometry cut designed to identify stars with proper motions and radial velocities consistent with previously known members $\mathrm{(\Delta_{PM} < 10 \, mas/yr, |\Delta RV| < 5.4 \, km/s)}$ followed by optical spectroscopy to confirm youth indicators (H$\alpha$, Li). Sources that displayed youth indicators but disparate radial velocities were considered prospective members. We include all confirmed and prospective low-mass ($>$ K5) members in their census. We made further constraints on this sample using the Bayesian moving-group classification software, \texttt{BANYAN $\Sigma$}\footnote{\url{http://www.exoplanetes.umontreal.ca/banyan/}} \citep[][]{banyan}. This program utilizes input astrometry data and kinematic distributions of bona fide members to determine the probability of an object belonging to a particular moving group.

Using the radial velocities from \cite{S2017}, and proper motion \& parallax information from \textit{Gaia} Data Release 2 \citep[\textit{Gaia};][]{gaia}, we determine the likelihood of each star’s membership to $\beta$PMG using BANYAN. Prospective members with likelihoods of 20\% or less were excluded from the final sample. Due to the sensitivity of the probabilities on the radial velocity measurements, confirmed members were kept regardless of the probability of membership due to historical confirmations \citep[e.g.,][]{Torres2008, Schlied2010}. For example, if a source returned a low probability of membership despite multiple independent confirmations, we retain it in our sample. These studies noted the M dwarf count in the original census is incomplete due to its inconsistency with initial mass function informed stellar multiplicity \citep[][]{S2017}. These cuts resulted in a sample of 146 sources.

\subsection{TESS Observations} \label{sec:Rihanna}

The \textit{Transiting Exoplanet Survey Satellite}’s (TESS; \citet{TESS}) near all-sky survey was designed to search for planets around bright stars. It generates short cadence (120 s), long baseline ($\sim$1 month), and high S/N photometric data with more sensitivity to redder objects ($\mathrm{0.6-1 \mu m}$) which makes its data accessible for our analysis. 
The high flare rate of young low-mass stars ($\mathrm{>10^{-2} \, d^{-1})}$; \citet[][]{G2020}) combined with the short-cadence time sampling ensures a wide distribution of flare light curve shapes and amplitudes are well-sampled. 

The observing pattern coupled with the large field-of-view ($24^{\circ} \times 96^{\circ}$) of TESS translates to adjacent sectors overlapping at high ecliptic latitudes during the primary mission. As a nearby moving group, the \bpmg members are not localized within a particular region of the sky thus some stars are observed for many consecutive sectors while others have none.

Along with the varied observation duration, other factors which can alter data extraction include multiple star systems and nearby field stars. Since TESS pixels are 21\arcsec $\times$ 21\arcsec, light from multiple sources within close proximity to another on the sky will blend together. Target pixels for each star were visually inspected with a \textit{Gaia} source overlay generated by \texttt{lightkurve}\footnote{\url{https://docs.lightkurve.org/}} \citep[][]{lightkurve} with close companion information provided by \citet{S2017}. We note that older faint field stars are unlikely to contribute significantly to any observed variability \citep[][]{G2020, Ilin21}. As a result, $\beta$PMG members are kept in our sample if they are brighter than field companions by one magnitude in the \textit{Gaia} G band. Any $\beta$PMG members with  $\beta$PMG companions were to be discarded due to flare source ambiguity. Late K dwarfs ($>$ K5) are included in this sample as a transitional boundary into the M dwarf regime due to the intrinsic uncertainty in determining the physical parameters of young, low mass stars. 

All light curves used were extracted from target pixel files calibrated by the Science Processing Operations Center (SPOC; \citet{jenkins+16}). We use  \texttt{lightkurve} to query and investigate the two-minute cadence data for every sector available for each source. While 20-second cadence observations are available for a subset of the sources in our sample, we opt to only use two-minute cadence due to the enhanced complexity in flare extraction at a higher cadence and for ease of comparison between data sets.

\subsection{Rotation Rates}

\begin{figure*}
	\includegraphics[width=2\columnwidth]{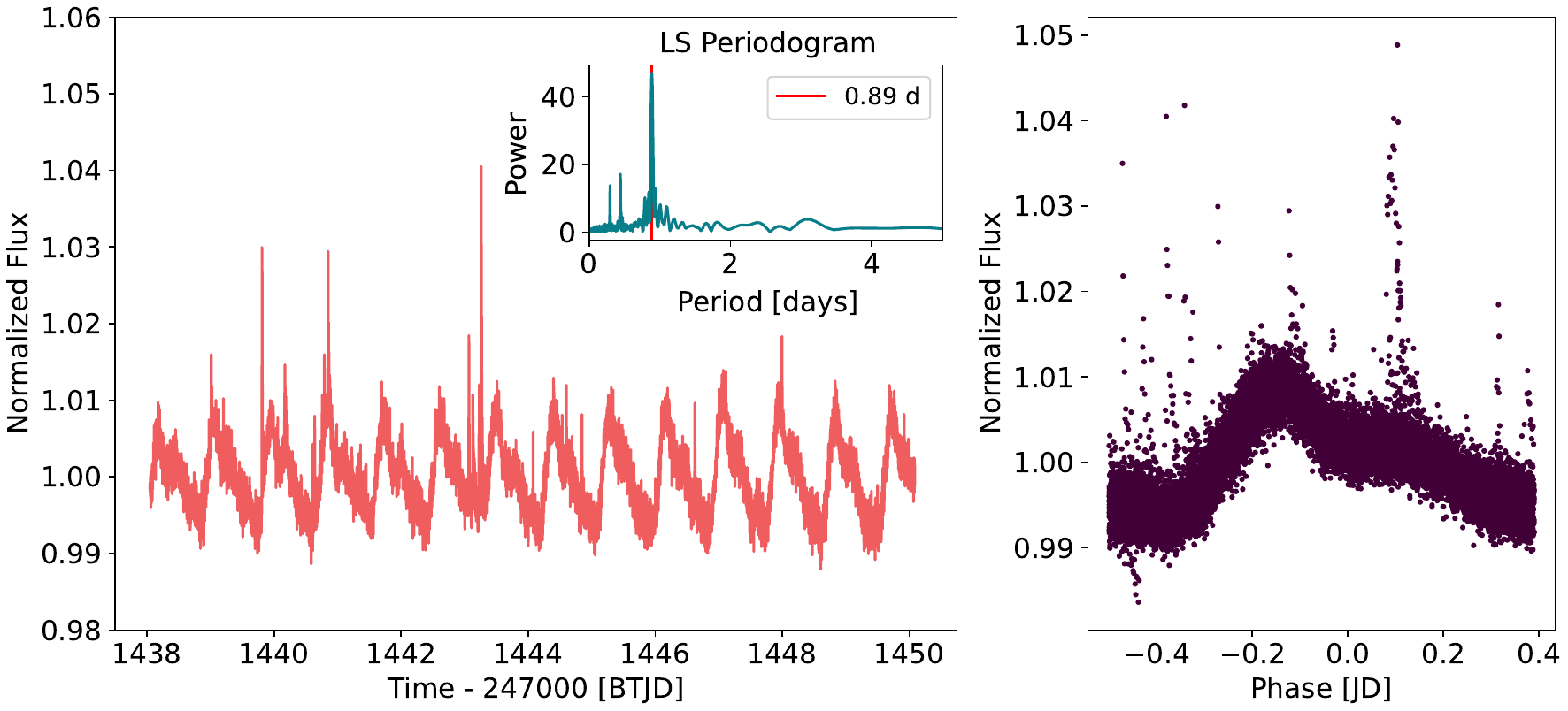}
    \caption{All rotation periods within this sample were derived from a sector of TESS observations. We use each light curve (left) with the power spectrum from a Lomb-Scargle periodogram (inset). The maximum power returned is used to phase-fold the light curve (right).}
    \label{fig:lsmethods}
\end{figure*}

The strength and prevalence of the magnetic field of a star are linked to its rotation rate and properties of its interior convection \citep[][]{Wright2011}. Determining the rotation period for active low mass stars is straightforward. These stars, due to their enhanced magnetic activity, often have distinct star spot patterns which can evolve slowly over time. The strong and regular variability (Figure \ref{fig:spotmod}) allows for M dwarf periods to be fairly well determined with a simple Lomb-Scargle periodogram. For less active or more distant stars (Fig. \ref{fig:spotmod}, bottom), this method is not as successful due to the absence of visible periodic variability and diminished flux.  Period searches for single-sector observations are only sensitive to periods $<$ 10 days due to statistical constraints. Since these stars are young as well as low mass, they can be expected to be fast rotators so this does not limit accurate period determination. All periods found for this sample were phase-folded and individually confirmed  to have a significance of more than 10\% compared to any secondary peak in the power spectrum (Fig. \ref{fig:lsmethods}) Any sources which did not meet this criteria were excluded.

Our membership confirmation, target pixel analysis, and period calculation generated a sample of 49 K and M dwarfs (Table \ref{table:theone}). The demographics of this sample after quality cuts are not fully representative of the expected initial mass function shape with the ultra-low mass regime ($>$ M5) poorly populated (6\% of final sample). 

\begin{deluxetable}{ccc}
\tablewidth{0pt}
\tablecaption{$\beta$ Pictoris Low-mass Stellar Sample}
\tablehead{
\colhead{Column} & \colhead{Units} & \colhead{Description}
}
\startdata
1 & -- & TESS Input Catalog Number \\
2 & deg & Right Ascension \\
3 & deg & Declination \\
4 & -- & Spectral Type \\
5 & day & Photometric Rotation Rate \\
6 & erg & TESS band Luminosity ($\mathrm{L_{\textit{TESS}}}$) \\
7 & -- & Fitted $\alpha$ \\
8 & -- & Fitted $\alpha$ (error) \\
9 & -- & Fitted $\beta$ \\
10 & -- & Fitted $\beta$ (error) \\
11 & $\mathrm{d^{-1}}$ & $\mathrm{R_{31.5}}$ \\
12 & $\mathrm{d^{-1}}$ & $\mathrm{R_{31.5}}$ (error)\\
13 & s &  Min. Equiv. Duration \\
14 & s &  Med. Equiv. Duration \\
15 & s &  Max. Equiv. Duration \\
\enddata

\tablecomments{All members of this sample came from the \citet{S2017} sample. Spectral Type references can be found therein. TIC Number, RA, Dec and $\mathrm{L_{TESS}}$ derived from the TESS Input Catalog \citep{https://doi.org/10.17909/fwdt-2x66}.}
\label{table:theone}
\end{deluxetable}

\subsection{Modular Flare Detection} \label{sec:marigolds}
With the sources and respective light curves quality-vetted, we move on to flare detection. In optical light curves, a classical flare has a sharp increase in brightness followed by a gradual decay phase with a total duration spanning from minutes to several hours. \cite{dave14} produced a morphology model for optical flares using \textit{Kepler} which captures these features as a sharp rise with an exponential decay. Most flares can be characterized with this model but more detailed studies into flare morphology show more diversity in various band passes \citep[][]{mac21,paudel21}, with sympathetic (induced) and superimposed flaring \citep[][]{G2020}, and quasi-periodic pulsations in the decay branch \citep[][]{N+04, P+16, J+19}. Since the energy detected from a flare is a function of the light curve morphology and the assumed power spectrum, these two factors can drastically alter the total energy extrapolated. Our methods use a blend of canonical and machine learning techniques to increase our ability to accurately extract flare morphology.

Most flare finding algorithms rely on iterative detrending of the light curve to remove periodic variability followed by an outlier cut. We utilize this method using the software \texttt{AltaiPony}\footnote{\url{https://github.com/ekaterinailin/AltaiPony}} \citep[][]{Ilin21}, informed with a convolutional neural network (CNN) trained to find flares, \texttt{stella}\footnote{\url{https://adina.feinste.in/stella/}} \citep[][]{stella}. \texttt{AltaiPony} is the main software used for this analysis -- capable of query, detrending, and characterization of flares for the entire sample. Before launching into analysis, however, we ensure that \texttt{AltaiPony} finds flares efficiently given our sample. For \texttt{stella}, the neural network can be given any light curve and predict the likelihood that a feature is a flare. The original calibration set of flares was built from a robust ($>$2000) sample of M dwarf stars observed with TESS. By ensembling a subset of the flare models generated from that analysis, \texttt{stella} returns the likelihood that a given feature is a flare. Since detrending can remove smaller features, training for shape on raw light curves rather than outliers makes \texttt{stella} particularly sensitive to the smaller flares in our sample.

\begin{figure}
	\includegraphics[width=\columnwidth]{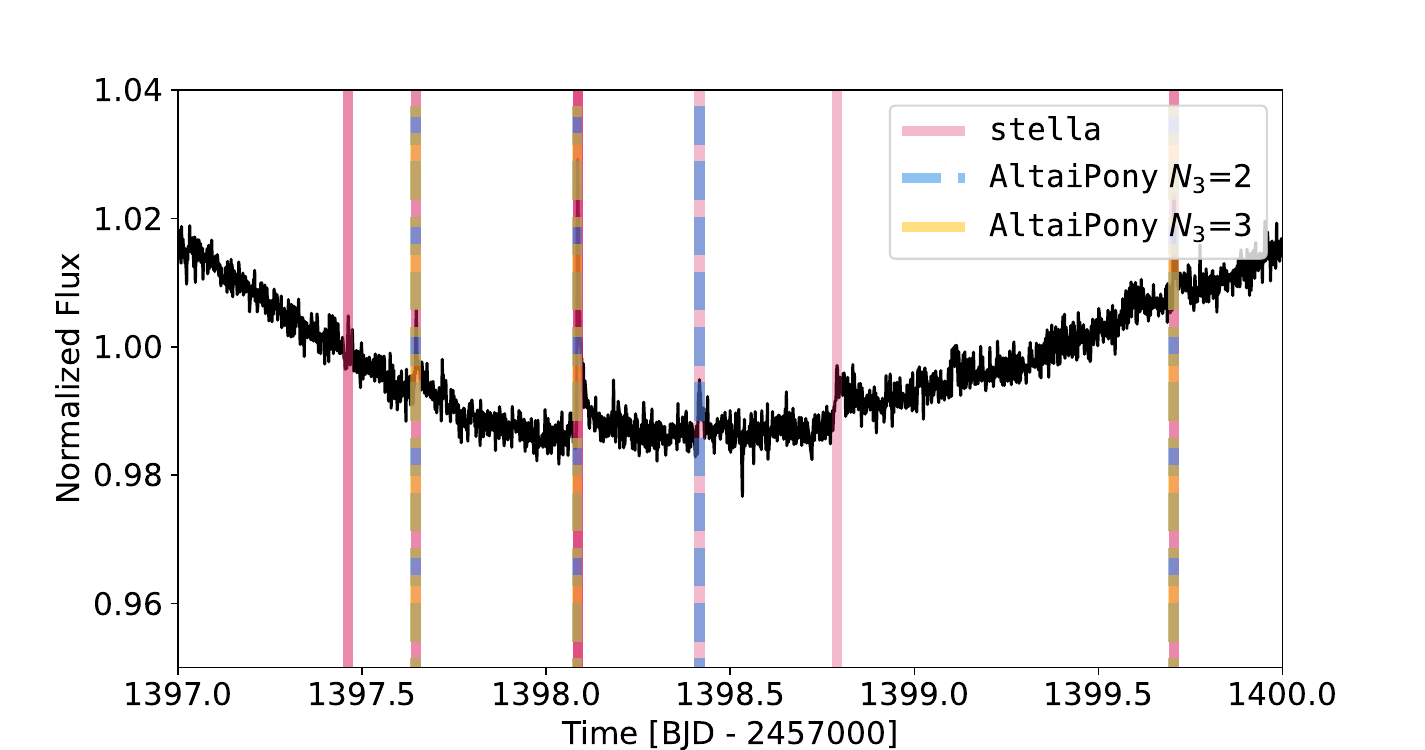}
    \caption{The flare finding results for \texttt{stella} (red) and \texttt{AltaiPony} ($N_3 = 2,\, 3$; blue, yellow) mark where each program defines the peak of a flare within a light curve excerpt for TIC 268637115. Without any adjustments to the standard number of points above the baseline, the number and characteristics of the flares found overlap about half the time. Tuning the \texttt{AltaiPony} parameters leads to better agreement. The flare near 1398.5 MBJD is only recovered by \texttt{AltaiPony} with $N_3 = 2$ and \texttt{stella}. Since \texttt{AltaiPony} relies on an outlier cut, it reliably finds large flares while \texttt{stella} focuses on finding a characteristic shape. A closer look at the flagged \texttt{stella} times shows more non-classical flares. By tuning to maximize their agreement on visually confirmed flares, we determine the optimal \texttt{AltaiPony} parameters for this dataset.}
    \label{fig:altai.vs.stella}
\end{figure}

\texttt{AltaiPony} relies on a Savitzky-Golay \citep[][]{SavGo64} filter to remove periodic signals after which it will flag any signal above the median that fits within a few modular parameters including outlier cuts (N1) and points above the baseline (N3) (defaults; $N_1 = 3\sigma$, $N_3 = 3$). We run the default parameters for both \texttt{AltaiPony} and \texttt{stella} on our entire sample and manually inspect the results between the two to determine the optimized  $N_1$ and $N_3$ parameters for \texttt{AltaiPony} which maximize the real flares recovered (Fig. \ref{fig:altai.vs.stella}). Through this process, we keep the $3\sigma$ outlier cut but move from three to two points above the baseline. This shift is necessary given that for smaller flares, most of the exponential decay phase is likely to be obfuscated by the local scatter. Therefore, finding these flares depends mostly on their peaks. These adjustments allow for improved identification of small amplitude and short duration flares while minimizing false detections. Our methods for accommodating the expected breakdown in flare detectability for more distant and later type M dwarfs is discussed in Section \ref{subsec:inj}.

\begin{figure}
	\includegraphics[width=\columnwidth]{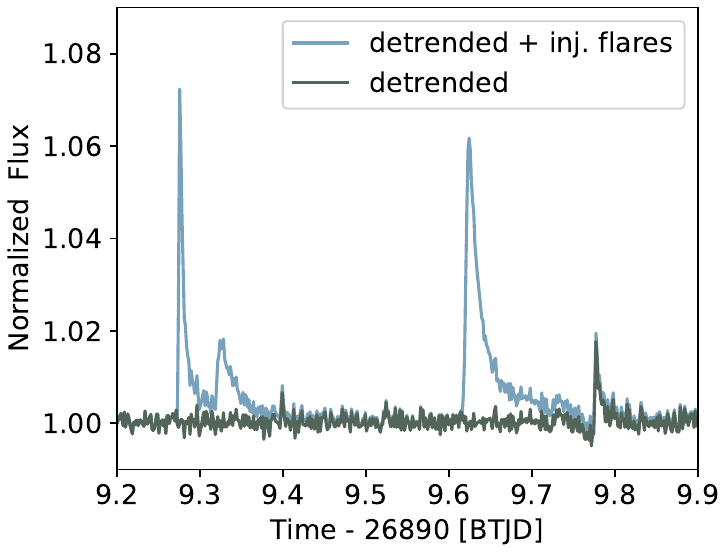}
    \caption{Flares injected by \texttt{AltaiPony} are shown in blue with a portion of the underlying light curve for TIC 269077806 shown in dark green. By running the flare finding algorithm on this data set, we can compare the characteristics of the injected flares to the recovered to develop scaling ratios. These ratios can then be used on the original extracted flares to scale them to their likely intrinsic distribution and shape.}
    \label{fig:inject}
\end{figure}

\subsection{Flare Modeling and Characterization Methods} \label{subsec:inj}

After identifying flares within a given light curve, key parameters of a flare (i.e, duration, full width half maximum) can be extracted by fitting an optical flare model. How well a model fits a certain flare is dependent on \texttt{AltaiPony}’s ability to find it within the light curve.
Finding low energy and low amplitude flares in a noisier or highly variable light curve is more difficult than finding high energy flares. We can quantify this effect using flare injection and recovery. By repeatedly injecting model-generated flares of various shapes into light curves and attempting to recover them, we can then draw relationships between the number and properties of injected versus recovered flares (Fig. \ref{fig:inject}). This provides a scaling ratio that can correct the real recovered flares to the parameters that are more intrinsic to their original morphology and distribution. We run this process for each light curve 100 times with injected flares spanning $10^{30}-10^{34}$ ergs at a chosen frequency of daily $[\mathrm{d}^{-1}]$. This ensures about 2500 synthetic flares are generated per light curve, effectively sampling each energy regime resulting in a well constrained and statistically dependable set of parameter scaling relations. 

The two main relations derived from this analysis are the recovery probability and the injected/recovered equivalent duration ratio. Recovery probability refers to the likelihood that a flare of a certain shape will be detected with \texttt{AltaiPony} which is a function of the peak amplitude and the duration. To determine this, a flare of a given energy will be injected into a light curve at some frequency. Given we know how many were injected and how many were recovered, the ratio between the two provides the recovery probability for a flare of that energy. Results from this analysis show a sharp recovery function near 60\% for all stars in the sample. Thus, flares with a recovery probability of $<60\%$ are discarded due to their characteristics likely being poorly constrained.

The equivalent duration (ED) refers to the duration of quiescent stellar emission which would equal the energy released by the flare. This is equivalent to integrating under the detected flare in the detrended light curve. 

\begin{equation}
    ED \, [s] = \int_{t_0}^{t_f} \frac{F_{flare} - F_{star} }{F_{star}} dt 
\end{equation}

The ED is particularly sensitive to the time at which the decay branch of a flare ends and thus the signal to noise. Since this metric translates to the energy released, its relationship to other stellar parameters and observed quantities offer insight into the physical mechanisms governing flares.

With a frequent and finely sampled injection parameter space, both these values are well approximated for each flare. We can calculate the unbiased EDs and flare occurrence rates by dividing the original detected values by the injection/recovery calibrated ratio $(\mathrm{X_{intrinsic} = X_{detected} / \,ratio})$.
Since the injection/recovery ratios depend on the signal to noise and variability, each sector's light curve is treated individually and flares detected for a single source observed across multiple sectors are compiled in postprocessing. With an understanding of the intrinsic flare distribution, we can compare different metrics for the flaring activity to the physical parameters of the star.

\subsection{Flaring Activity Summaries and Flare Energies}

There are two main methods for summarizing the flaring activity for individual stars: the fractional flare luminosity and the cumulative flare frequency distribution. The fractional flare luminosity (FA) directly compares the total flux released by the flares to the luminosity of the star averaged over the duration of the observation. This is often shown as $\log_{10}(\mathrm{FA})$ due to the smaller bolometric contribution. While this is the most direct comparison, this method is weighed heavily by robust flare detection and good resolution of total flare morphology. This can become particularly important for M dwarfs that are more distant and fainter. The exponential tail of these flares can be lost in the intrinsic light curve scatter as well as the two minute cadence of TESS observations, effectively reducing EDs and the resulting FA.

The second method uses the total observation time to calculate flaring rates at a given ED or higher known as a cumulative flare frequency distribution (FFD). The FFDs generated grant an opportunity to study the distribution of flares energies from a given source and how it varies between sources. This distribution is often fit with a cumulative power law taking the form

\begin{equation} \label{eq:CPL}
     \nu = \frac{\beta}{|\alpha - 1|} \times E^{-\alpha + 1},
\end{equation}

where $\alpha$ and $\beta$ are a power index and normalization factor, $\nu$ is the frequency of flares in a chosen unit (days), and $E$ is either the TESS band energy of the flare in ergs or the ED. Since EDs are directly related to energy, we convert from equivalent duration to total energy released. As mentioned previously, the wavelength dependence of the emission spectrum is not fully constrained and can vary between flares. Ignoring notable spectral features, the TESS bandpass captures less than 20\% of the total emission from a 9 kK blackbody.

Without spectral and flare temperature constraints, these fractions can only approximately describe the total emission. To maintain the integrity of this sample, all energies within this work will be computed in the TESS band where the flares were observed. We calculate the energy of each flare using the TESS band magnitude, $T_{\mathrm{mag}}$, the distance to the star, and the equivalent duration. Distances for each star were calculated using the \textit{Gaia} DR2 trigonometric parallax. The total energy in the TESS band released by the flare is the quiescent luminosity in the TESS band multiplied by the equivalent duration. \citet{sullivan15} derive the zero-point magnitude flux in the TESS band $(F_0 = 4.03 \times 10^6$ $ \mathrm{ergs}\, \mathrm{cm}^{-2} \, \mathrm{s}^{-1})$,

\begin{equation}
     F_{\mathrm{TESS}} = F_0 \times 10^{-T_{\mathrm{mag}}/ 2.5}.
\end{equation}

We use this conversion with the distances, $d$ in cm to calculate the energy of each flare, assuming no extinction due to their proximity,

\begin{equation}
    E_{\mathrm{flare,TESS}} = \mathrm{ED} \times 4 \pi d^{2} \times F_{\mathrm{TESS}}. 
\end{equation}

With the calculated energies, we fit a cumulative power law with Eq. \ref{eq:CPL} and compute the rate of flares above $10^{31.5}$ ergs (near the average flare energy of the total sample).


\section{Analysis Results} \label{sec:results}

\begin{figure}
    \centering
    \includegraphics[width=\columnwidth]{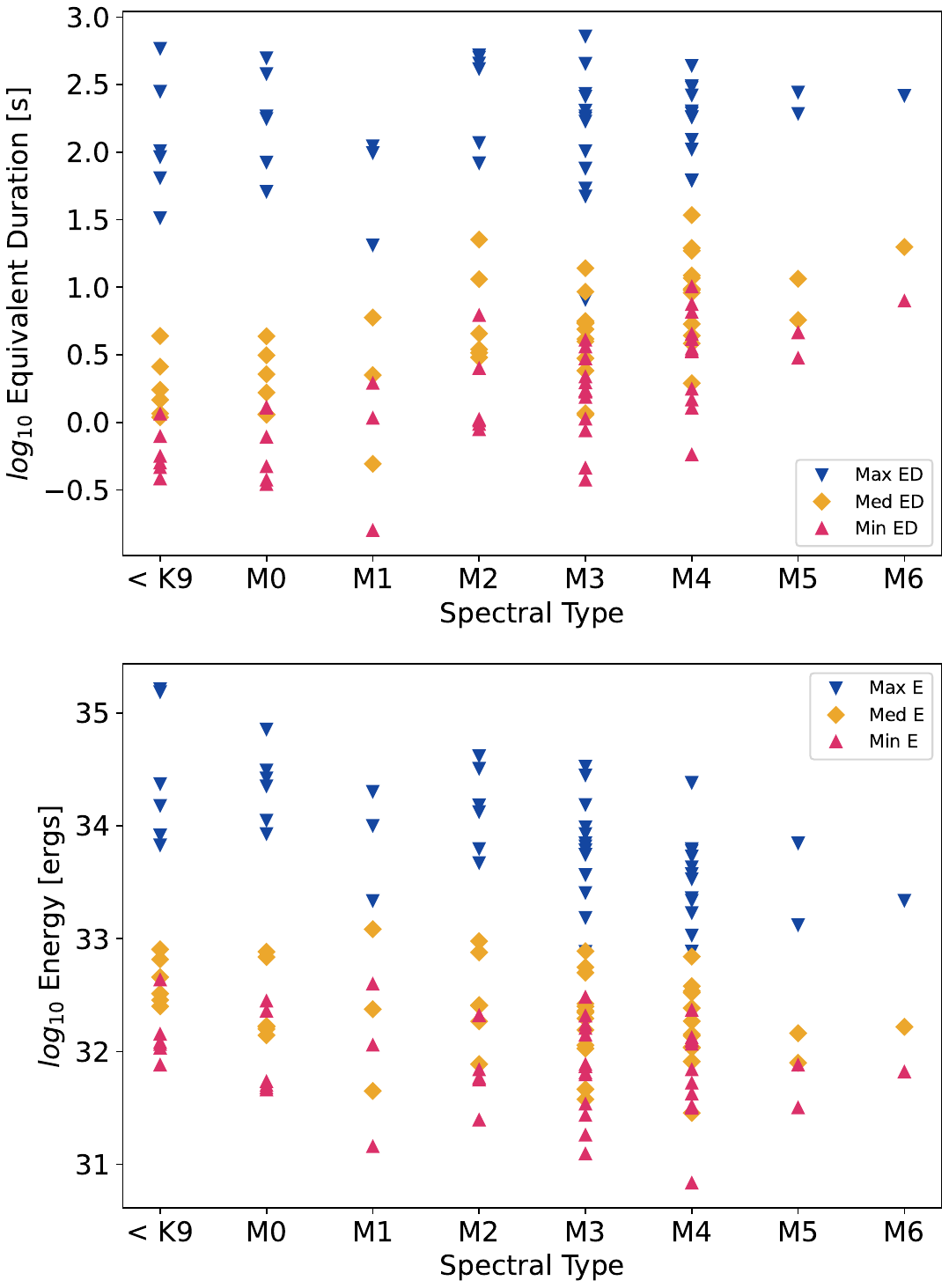}
    \caption{The maximum (blue), median (yellow), and minimum (magenta) energy flares detected for each star are listed with their spectral type in equivalent duration (top) and TESS band energy (bottom) space. In energy space, the median energy of flares remains flat with later spectral type. This also implies an intrinsic connection between the mechanism generating flares and the mass of the star. This is accompanied by a decreased effective temperature so while the total energy decreases, the comparative UV flux delivery compared to the blackbody of the star increases. Another notable feature is the flat maximum equivalent durations and accompanying downward shift in energy space as an anticipated change with decreasing luminosities.}  
    \label{fig:EDenergy}
\end{figure}
 
Figure \ref{fig:EDenergy} summarizes the flare morphology as a function of spectral type. For each spectral type, the maximum, median, and minimum ED (top panel) and TESS band energy (bottom panel) flares are plotted for each star.  The observed upward trend in minimum ED is a product of the detectability limit of TESS and \texttt{AltaiPony}. With increasing distance and decreasing brightness, the smallest flare we are able to detect decreases because their peaks are less distinguishable given the cadence of TESS and photometric scatter. Comparison of the largest flares is difficult due to their infrequent occurrence in a typical TESS sector duration of 28 days. Despite this, some of the largest EDs for this sample lie between a few seconds to minutes in the TESS band. The maximum energy flare detected decreases toward late-type stars as a result of their lower quiescent brightness. This is consistent with the flatter ED distribution given the decreasing luminosity of later-type stars. We see a similar effect where the detection limits push the median ED upwards, leveling out the medians in energy space.
 
The decreasing stellar effective temperature translates to the fractional UV contribution from flares increasing. So, while the total energy released by large flares decreases with spectral type, the impact of flares on planetary atmospheric composition and structure is more significant for young later-type stars. In practical terms, the median energy for flares for the $\beta$PMG M dwarfs tends to lie between $10^{32} - 10^{33}$ ergs in the TESS band. This sample confirms previous understanding \citep[][]{dave+19, Medina20, fein+20, Ilin21} that with later type stars,  flares with energies that fall within this range are relatively common and our flare detection methods are readily capable of finding them for all spectral types in our sample.

\begin{figure}
    \centering
	\includegraphics[width=\columnwidth]{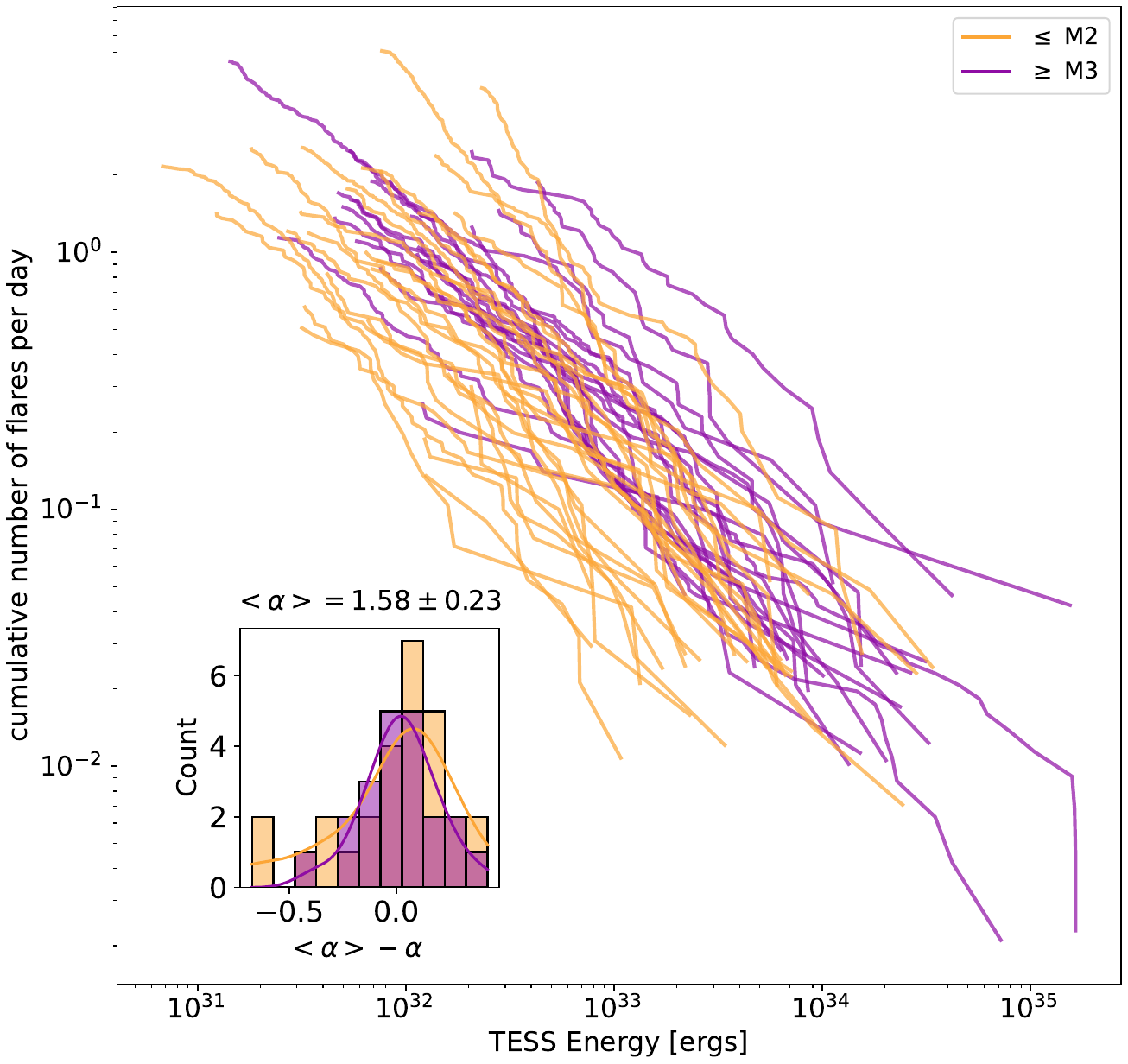}
    \caption{Flare Frequency distributions for $\beta$ Pictoris low-mass stellar sample split between early (orange) and mid/late- (purple) type stars. The denser sampling close to the lower energy regime is indicative of the overall prevalence of low energy flares compared to those of high energy. The average power law for this sample is $\alpha = 1.58 \pm 0.23$ with a similar overlap for both populations. The inset shows a non-negligible difference between FFDs though the distribution has relative normal shape. This is due to intrinsic variability of activity in our sample driven by differences in spindown timescales and internal dynamos across the low mass regime. All FFDs occupy a similar parameter space with early types showing a broader spread to smaller flares for similar rates. In addition to these natural variations, some FFDs are not fit well by a single power law due to their morphology. Of note, the largest flares in this sample are mostly from later type stars.} 
    \label{fig:spagetti}
\end{figure}

Figure \ref{fig:spagetti} shows the FFDs for every star in the $\beta$PMG low mass stellar sample split between early and later type stars. The break at M2 is motivated by the partial to fully convective boundary. On average, both early- and late-type stars occupy the upper-right portion of the FFD diagram, consistent with those stars producing more and higher energy flares. Comparatively, the lower-left portion is mainly occupied by earlier-type stars. In another way, for a given energy, earlier-type stars can flare similarly or less often than their later-type counterparts. This wide range of activity level is a direct consequence of the varied spin-down rate, with earlier types more likely to evolve sooner or more rapidly \citep{Ilin21}.

Each FFD's power law was fit using Eq. \ref{eq:CPL} resulting in $\alpha$ values ranging from 1.15 to 2.25, with an average of $1.58\, \pm \, 0.23$. In addition to the qualitative FFD distribution, the fitted $\alpha$ values are broadly consistent within the spectral type split. Due to the nature of power law fits and occurrence statistics, it can be difficult to directly compare slope values. \cite{Ilin2019} and \citet{Medina20} find the average slope of older M dwarfs in open clusters and local field M dwarfs to be consistent with 2, with larger variations for younger stars. We also observe a large variation in FFD slopes here, but the average slope is significantly shallower. \citet{dave+19} also saw a tentative shift to shallower slopes with later types but lacked the population to make a conclusive statement. \citet{Feinstein+2024} also find similar shallow slopes across thousands of young stars less than 300 Myr with no discernible trends with spectral type. Flares are expected to continue to lower energies but these are beyond the detectability of TESS. Previous studies have explored the potential of a broken power law fit to account for this variation \citep[][]{paudel21}. We do not explore this option to maintain integrity for comparison with larger-scale flaring studies.

\begin{figure}
        \centering
	\includegraphics[width=\columnwidth]{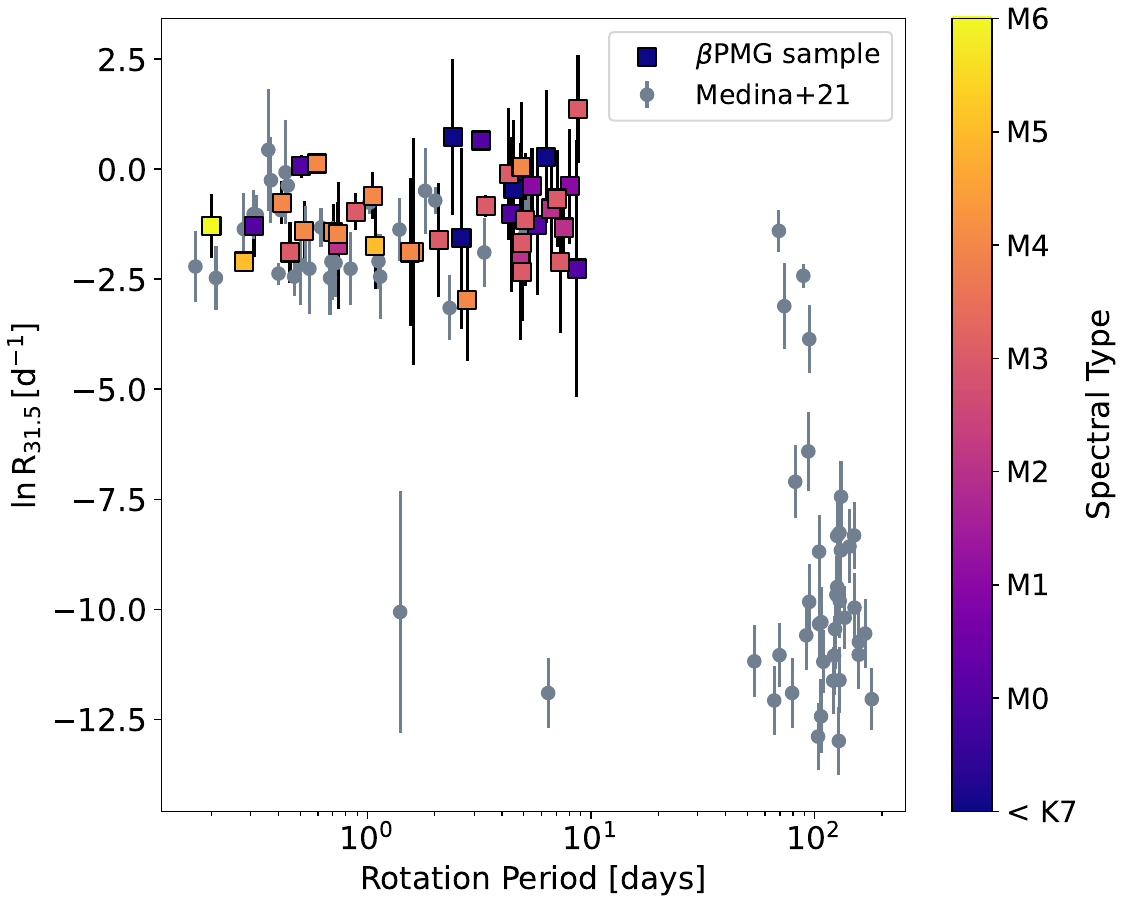}
    \caption{The rate of flares above $10^{31.5}$ ergs compared to the stellar rotation period. The rates derived in this sample are shown in color as a function of spectral type. The grey points plotted underneath are the results from \citet{Medina20}. The flaring rates found for this sample are systematically consistent with the field M dwarf flaring rates. The $\beta$MPG sample also occupies the `saturated' regime where with decreasing rotation period, the flaring rate remains fairly constant. Due to the fine sampling of flare injection, the recovery probability of the flares in this sample is well determined.}
    \label{fig:medina}
\end{figure}

Figure \ref{fig:medina} shows the flaring rate of the $\beta$PMG sample as a function of rotation period. The color bar provided shows our sample as a function of spectral type. Under-plotted here are the results from the \citet{Medina20} survey of a volume complete sample of M dwarfs within 15 pc. While their study was mainly focused on determining a relationship between spectroscopic activity indicators and flaring activity, they also derived rotation rates and FFDs for each of their stars. Given the volumetric range, a subset of their sample overlaps with the \bpmg. There is a distinct gap between the fast and slow rotators. With the $\beta$PMG sample concentrated in the fast rotator regime, the reorganization of the stellar magnetic field required to induce spin-down does not occur within the age of the $\beta$PMG \citep{Ilin2019}. There is no strong dependence of the rotation period on spectral type. 
With respect to the flaring rate, $R_{31.5}$ represents the rate of flares with an energy of $3.16 \times 10^{31}$ ergs in the TESS band. This rate is defined by \citet{Medina20} as the rate that offers a summary of flare activity for all flares detected with TESS. For a more direct comparison to previous work, we examine the optical flare rates measured in this study for the benchmark planet host AU Mic. We find they lie between the values found in \citet{G+22} and \citet{Feinstein+2024} for the same two observations. The differences are likely due to the different flare identification and modeling approaches, and in particular the confidence constraints and greater detection sensitivity enabled by machine-learning techniques \citep[e.g.][]{Feinstein+2024}.
With the fitted power law relations, the \bpmg exists in the `saturated' regime. Since the rotation and flaring rates are intrinsically tied to the magnetic dynamo, the saturation in the flaring rate offers insight into the mechanisms related to generating flares. Whichever process is responsible for determining how the energy is released from flares does so such that the total flaring rate is similar for stars independent of mass. Compared to the \citet{Medina20} sample, the \bpmg sample overlaps their rapid rotator sample consistently. As noted in the FFD analysis, the smaller $\alpha$ distribution in the \bpmg sample suggest shallower power law relations for younger stars which aligns with previous studies \citep[e.g.][]{dave+19,Jackman+2021, fein+20, Feinstein+2024}. Since the $R_{31.5}$ values overlap with the older rapid rotators, the \bpmg sample is likely producing more large flares overall. Since the rotation rates are comparable, this may be due to the inflated radii from their pre-main sequence state allowing for a favored relationship between their magnetic filling factors and available surface area \citep[][]{Benz2010}. Ongoing multiwavelength flare campaigns (particularly in the X-ray) \citep[][]{paudel21,mac21,tristan+23}, solar-analog comparisons \citep[][]{Benz2010, Howard+22b, name+23}, and theoretical modeling will offer insight into physical mechanisms responsible for energy partition and release as a function of low-mass star age.

\section{Discussion \& Conclusions}\label{sec:cons/disc}

We characterize the flaring activity of 49 K and M dwarf members of the \bpmg from their respective TESS 2-minute cadence observations. To maximally extract flares, we use a combination of \texttt{AltaiPony} and \texttt{stella} to calibrate a flare finding program sensitive to low energy and low amplitude flares common for M dwarfs. Analysis of the results reveals three major conclusions.

For the young $\beta$MPG, the FFDs of different spectral types occupy a broadly similar parameter space but there are discernible trends. Within this sample, earlier type stars can flare at similar or lower rates for a given energy compared to later type stars. This is likely due to the difference in spin down rates between early- and later-type stars in this sample \citep{Ilin21}. Our analysis shows distinct behaviors within the population including a cumulative FFD average slope of $1.58\, \pm \, 0.23$. This is shallower than observed in older low-mass star samples and corresponds to more large flares overall and has been seen in other samples \citep{dave+19, Jackman+2021, Feinstein+2024}. The stars are also all rapid rotators and occupy the ``saturated" flaring regime observed in \citet{Medina20}.

Constrained optical flaring rates offer insight into the physical mechanisms governing flares including generation and spectral energy partition. The flatter distribution in maximum equivalent durations (Fig. \ref{fig:EDenergy}) offers insight into the types of events that can be expected from these stars. The bulk flare behavior of low-mass stars in the \bpmg is largely consistent with the growing cache of other young associations studied with TESS and other large optical flare analyses \citep{dave+19,G2020, Ilin2019, Ilin21, Feinstein+2024}. In addition, the evidence toward population differences between low-mass spectral types becomes more apparent in the analyses presented here.
This motivates continued observations and investigations to further represent and understand optical flare statistics and evolution for late-type stars.

With respect to exoplanet atmospheres, the UV regime is the driving force of photochemistry. Since the spectral energy distribution and flare temperatures are poorly constrained, a blanket blackbody assumption is likely to introduce errors on energy delivery upwards of tens of percent in the optical and by an order of magnitude in UV emission \citep[e.g.][]{K2013, Ward2020,brasseur2023,jackman2023}. Several methods are currently being explored to combat this. One method is by studying quasi-periodic pulsations (QPP) which can be observed in TESS light curves. These oscillatory waves travel along the magnetic field lines of the star where frequencies at maximum amplitude are linked to their location in the star's atmosphere \citep[][]{N+04}, offering context to flare structure. Another method is through direct multiwavelength observations \citep{mac21, paudel21,tristan+23,paudel2024}. By better understanding the relationship between bandpasses, we can further evaluate energy partition and emission mechanisms and timing. Observations in the UV require space-based observatories while the optical is accessible from both space and the ground. Measured optical flaring rates coupled with these methods will provide more accurate inputs for atmospheric modeling and interpretation.

The flare energies of low-mass stars in the \bpmg have a median value consistent with optical Carrington-like events occurring near daily. Regarding the Sun, the Carrington event was the strongest solar flare ever recorded, releasing between $10^{32}$ and $10^{33}$ ergs \citep[][]{C1859}. Within the context of habitability, the excess luminosity provided by flares increases the total flux at the boundaries of the habitable zone. With the low effective temperatures of low mass stars, the habitable zone also exists at close radii, thus increasing the likelihood of direct flare-planet interactions. In addition to this, the pre-main sequence evolution of M dwarfs can spend billions of years at luminosities notably larger than its main sequence luminosity \citep[][]{Baraffe2015}. These factors together can render developing planets incapable of sustaining an atmosphere or appreciable water fraction \citep[][]{T2019}. The atmospheres that survive this pre-main sequence phase are then exposed to the excess UV and X-ray emission and associated coronal mass ejections from flares. Quantifying how the flaring rate changes with age can place limits on the expected conditions of a planet within the habitable zone. Estimates of the surface or atmospheric conditions can inform and interpret the observations necessary to better characterize a planet and help better define the co-evolution of stars and planets \citep{Rug2015}.

Moving forward, the development and use of the flare pipeline presented here will allow us to extend this analysis to the low-mass members of other moving groups, and thus other age-distinct samples. With a larger sample of low-mass stellar flare properties and rates spanning a wide range of ages, we plan to expand this work and investigate the evolution of flare activity during critical times in the co-evolution of stars and planets. We also plan to investigate the dependence of other properties such as flare morphology and star spot coverage on age and stellar parameters.

This material is based upon work supported by NASA under award number 80GSFC21M0002. JE thanks the LSSTC Data Science Fellowship Program, which is funded by LSSTC, NSF Cybertraining Grant \#1829740, the Brinson Foundation, and the Moore Foundation; their participation in the program has benefited this work. JE and EG's research was also supported by NASA’s Astrophysics Data Analysis Program through grant 20-ADAP20-0016 and carried out in part at the Jet Propulsion Laboratory, California Institute of Technology, under a contract with the National Aeronautics and Space Administration (80NM0018D0004).
This paper includes data collected by the TESS mission and the Tess Input Catalog (TIC), which are publicly available from the Mikulski Archive for Space Telescopes (MAST): 10.17909/fwdt-2x66. This work has made use of data from the European Space Agency (ESA) mission Gaia, processed by the Gaia Data Processing and Analysis Consortium (DPAC). Funding for the DPAC has been provided by national institutions, in particular the institutions participating in the Gaia Multilateral Agreement.


\bibliography{project}{}
\bibliographystyle{aasjournal}

\end{document}